\documentclass[aps,pre,superscriptaddress,onecolumn]{revtex4}

\usepackage{graphicx}
\usepackage{dcolumn}
\usepackage{bm}
\usepackage{amsfonts}
\usepackage{amsmath}

\begin{document}


\title{Towards a data-driven characterization of behavioral changes induced by the seasonal flu}

\author{Nicol\`o Gozzi}
 \email{N.Gozzi@gre.ac.uk}
 \affiliation{Networks and Urban Systems Centre, University of Greenwich, London, UK}

\author{Daniela Perrotta}
 \affiliation{Max Planck Institute for Demographic Research, Rostock, Germany}
 
\author{Daniela Paolotti}
 \affiliation{ISI Foundation, Turin, Italy}
 
\author{Nicola Perra}
 \affiliation{Networks and Urban Systems Centre, University of Greenwich, London, UK}
  \affiliation{ISI Foundation, Turin, Italy}

\begin{abstract}
In this work, we aim to determine the main factors driving behavioral change during the seasonal flu. To this end, we analyze a unique dataset comprised of $599$ surveys completed by $434$ Italian users of Influweb, a Web platform for participatory surveillance, during the $2017-18$ and $2018-19$ seasons. The data provide socio-demographic information, level of concerns about the flu, past experience with illnesses, and the type of behavioral changes implemented by each participant. We describe each response with a set of features and divide them in three target categories. These describe those that report i) no  ($26\%$), ii) only moderately ($36\%$), iii) significant ($38\%$) changes in behaviors. In these settings, we adopt machine learning algorithms to investigate the extent to which target variables can be predicted by looking only at the set of features. Notably, $66\%$ of the samples in the category describing more significant changes in behaviors are correctly classified through Gradient Boosted Trees. Furthermore, we investigate the importance of each feature in the classification task and uncover complex relationships between individuals' characteristics and their attitude towards behavioral change. We find that intensity, recency of past illnesses, perceived susceptibility to and perceived severity of an infection are the most significant features in the classification task. Interestingly, the last two match the theoretical constructs suggested by the Health-Belief Model. Overall, the research contributes to the small set of empirical studies devoted to the data-driven characterization of behavioral changes induced by infectious diseases.
\end{abstract}

\maketitle

\section{Introduction}
Understanding and influencing behavioral change are key challenges for a range of disciplines such as Medicine, Psychology, Epidemiology, Social Policy and Computational Social Science. Most of the relevant literature focuses on \emph{complex interventions} designed to nudge populations to adopt healthier and safer habits. Recommendations about the increase of physical activity~\cite{michie2009effective}, quit smoking~\cite{west2010behavior}, change of behaviors in workplaces~\cite{ivers2012audit}, or promoting safe sexual behaviors~\cite{albarracin2005test} are classic examples. Yet, even in the absence of top-down (complex) interventions or external incentives, people may spontaneously modify their behaviors in response to different types of events. Infectious diseases are a notable example~\cite{funk,verelst,hk, swineflu, h1n1,kim2019incorporating}. Indeed, they may induce a range of (re)actions such as social distancing (e.g., reduction of contacts or mobility), the use of antivirals, change of diets and of personal hygiene practices~\cite{verelst}. Understanding these type of voluntary behavioral changes is crucial to i) improve our understanding of human dynamics under stress, ii) increase the predictive power and the realism of epidemic models, iii) improve communication campaigns from a public health perspective. In fact, human dynamics and human transmissible diseases are intertwined: an outbreak can induce behavioral responses which in turn can affect the course of the epidemic as a whole~\cite{nine,brauer2019challenges,meloni2011modeling,perra2011towards,poletti2009spontaneous,wang2015coupled,rizzo2014effect,poletti2012risk,zhang2014effects,moinet2018effect,abdulkareem2020risk,tyson2020timing}. While this observation is rather obvious, our understanding of how (and which) people change behaviors is extremely limited and largely anecdotal~\cite{verelst,nine,brauer2019challenges}. Arguably, the key issue is the lack of ground truth data. According to a recent review, only $15\%$ of the articles on the subject considered empirical data, most models being ``purely theoretical and lack(ing) representative data and a validation process''~\cite{verelst}. Nonetheless, relevant examples of studies that try to capture the behavioral dynamics coupled to epidemics using a data-driven approach can be found in literature. Surveys represent the most common data source for these research efforts~\cite{zhong, fierro, cohen, shim,kim2014community,rubin2014design}. Indeed, they allow to gather ground truth data, querying participants with specific questions about changes in behaviors, but they are limited by small sample sizes. An increasing number of works exploit social media data and other digital sources of information to characterize the behavioral response of humans to the spreading of infectious diseases~\cite{xia, springborn, pawelek, bayham,fast2015modelling,kim2019incorporating,fierro}. While this data allows to drastically increase the sample and to collect information in near real time (which is particular important during the unfolding of an outbreak), the ground truth is typically missing. Thus, a set of assumptions are needed to connect the online (i.e. people's posts) and offline worlds (i.e. behavioral changes). \\
In this context, we aim at advancing our comprehension of behavioral changes induced by infectious diseases with an approach that puts ground truth data about individuals' behaviors at its core. As an example of recurring and widely spread human transmittable disease, we consider the seasonal influenza. The reasons behind this choice are threefold. First, the World Health Organization, estimates the seasonal flu to result in about $3$ to $5$ million cases of severe illness, and about $290,000$ to $650,000$ respiratory deaths worldwide~\cite{who}. Besides the cost in terms of human lives, the seasonal flu represents also one of the main economic costs for public health systems~\cite{burden}.
Second, the few empirical studies on behavioral changes induced by infectious diseases are mostly focused on pandemics such as the $1918$ Spanish flu~\cite{hatchett2007public,bootsma2007public,markel2007nonpharmaceutical} and the $2009$ swine flu~\cite{swineflu,h1n1,rubin2014design,zhong,kim2019incorporating,kim2014community}. However, these events are quite rare; their timing, intensity, patterns, and media coverage are often out of the ordinary. The relevance of this (small) body of literature for other outbreaks and diseases is unclear. Third, we can leverage existing participatory Web platforms for digital surveillance of the seasonal flu to reach and query large numbers of users with explicit questions about behavioral changes. In fact, the yearly cadence of the seasonal flu allows the planning of regular data collection campaigns that can go beyond gathering information about disease's prevalence.\\

In this work, we combine health and behavioral data, collected from Web users, with a machine learning pipeline to characterize behavioral changes during the seasonal flu. In particular, we use Influweb~\cite{influweb1,influweb2}, a digital surveillance platform that since $2008$ collects data about the progression of the seasonal flu in Italy, to collect socio-demographic indicators, medical history of individuals, information regarding feelings, concerns towards the flu and to query users about changes in their behaviors induced by the disease. By studying the responses, we identify three classes of behavioral changes describing those that report i) no  ($26\%$), ii) only moderately ($36\%$), iii) significant ($38\%$) changes in behaviors. From this standpoint, we adopt a range of machine learning algorithms such as Gradient Boosted Trees (GBT)~\cite{greedy}, Support Vector Machine (SVM)~\cite{svm}, Logistic Regression (LG)~\cite{logisticregression} and Random Forest (RF)~\cite{randomforest} to solve a classification task in which a set of $23$ features (obtained from the responses and the characteristics of the epidemic) are used to predict the class of behavioral change of each user. In order to interpret the outcomes of the classifiers, we use SHapley Additive exPlanations (SHAP) values~\cite{shap1,shap2,shap3}. These allow to measure the importance of each feature in the classification task. Interestingly, we find  that GBT is able to correctly classify $66\%$ of the samples describing significant changes. Furthermore, we find that the severity and recency of past illnesses, the perceived susceptibility to the disease, the perceived severity of infection event are key factors driving behavioral changes. The last two drivers are in line with the constructs of the Health Belief Model (HBM)~\cite{hbm1,hbm2,hbm3}, which is by far the most commonly used psychological theory to explain and predict health-related behaviors. \\

Overall, these results quantify the extent to which individuals change behaviors
in response to the seasonal flu, uncover the key factors influencing
such changes, and quantify the limits of predictability of behavioral
classes in our sample. The research presented here contributes to the unfortunately still small set of empirical data-driven studies on disease outbreaks and behavioral changes. To the best of our knowledge, this is the first data-driven study focused on the seasonal flu and the first to use data on behavioral changes, induced by diseases, collected from a digital surveillance platform. The methodology and findings presented here pave the way to future extensions and generalizations able to capture multiple diseases, larger sample sizes as well as different countries. Finally, this methodology potentially represents a public health monitoring tool. In fact, the routine surveillance from Influweb are already communicated to the Italian National Institute of Public Health every week during the influenza season since $2015$. Additional quantitative assessment about the change in behaviors induced by the flu among the general population could represent a valuable insight for policy makers when communicating recommended behaviors to avoid contagion.

\section{Materials and methods}

\subsection*{Influweb dataset}
Influweb is a scientific project aimed at monitoring the activity of Influenza-like Illness in Italy with the aid of volunteers via the internet \cite{influweb1,influweb2}. It has been operational since $2008$ and it is part of the InfluenzaNet network, active in many other European and non-European countries  \cite{influenzanet1, influenzanet2}, such as The Netherlands, Belgium, Portugal, United Kingdom, Sweden, France, Spain, Denmark and Ireland. Throughout the years InfluenzaNet platforms  have been shown to be reliable sources of high-resolution and high-quality public health information~\cite{influweb-forecasting, influenzanet1, influenzanet5, influenzanet3}. In this work, we focus on the data collected through the Italian node of the platform, Influweb by means of three surveys: 
\begin{itemize}
\item \textit{Intake questionnaire}: is submitted when the user completes the registration and can be updated at the beginning of a new season; it covers demographic, geographic, socioeconomic (household size and composition, occupation, education, and transportation), and health (vaccination, diet, pregnancy, smoking, and underlying medical conditions) indicators.
\item \textit{Symptoms questionnaire}: is submitted weekly during the flu season. Participants are asked whether they experienced fever, respiratory or gastrointestinal symptoms (or "no symptoms") since their last survey. 
If symptoms are reported, further questions are asked to assess the syndrome (e.g, sudden onset of symptoms and body temperature).
\item \textit{Behavioral questionnaire}: is submitted during the flu season and contains questions related to perceptions towards the flu and behavioral attitude of participants.
\end{itemize}
Behavioral questionnaires have been introduced in Influweb since the $2017-18$ season, with the aim of shedding a light on behavioral aspects beyond the mere epidemiological data collection. A full account of the behavioral change questionnaire can be found on the Influweb Web page~\cite{influweb-behavioral-questionnaire}. In order to reduce the burden for the users, the behavioral questionnaires have been administered only during few weeks during each season, i.e. right before the peak and a couple of weeks after the peak. We refer the reader to the Supporting Information (SI) for more details. In total $599$ behavioral surveys were submitted by $N=434$ unique users: $73\%$ responded only once, $27\%$ instead more than once (in the same and/or across seasons). Consequently, while the large part of surveys are uniquely linked to a specific user, some are not. It is important to notice how sentiments, perceptions, and behaviors might vary during the flu season. Thus, we consider the $599$ surveys, rather than the users, the unit of analysis. As shown in the SI, this choice does not affect the overall results.

\subsubsection{Ethics statement}
Informed consent was obtained online from all participants enabling the collection, storage, and treatment of data, and their publication in anonymized, processed, and aggregated forms for scientific purposes. The Influweb website~\cite{influweb2} has a ``Privacy Policy" section in which the users who decide to enroll in the study can find all the information on who is responsible for the data acquisition and processing.



\subsection{Behavioral change classes}

\begin{figure}[t!]
  \centering
  \includegraphics[scale=0.2]{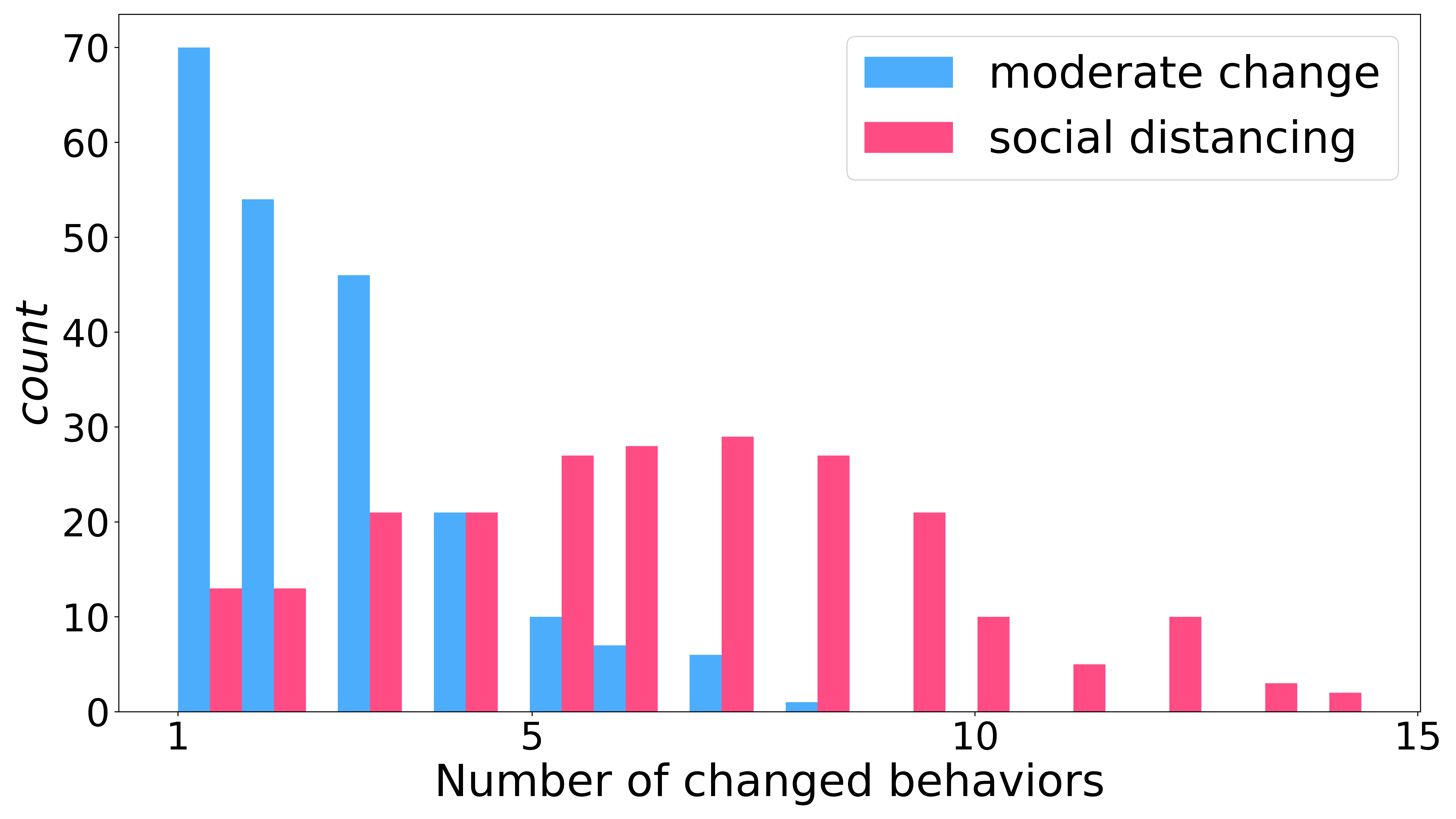}
  \caption{\textbf{Histograms of number of behaviors changed} for the class of \textit{social distancing} and for class of \textit{moderate change}. The two histograms look significantly different, with that of social distancing much more skewed towards higher values.}
  \label{fig:behaviors}
\end{figure}

In the behavioral survey, participants are asked, among other things, whether they have changed or not a number of behaviors in response to the flu. A natural categorization of the participants' responses would be to consider on one side individuals who do not change any of the possible behaviors, and on the other individuals who change at least one. However, our data suggest that this approach can be too restrictive. In fact, individuals who report to engage in behavioral change seem to form two different groups: 1) individuals who take only moderate preventive measures, such as more frequent hygiene measures, a healthier diet or the use of tissues when sneezing or coughing more often than usual, 2) individuals who, besides the previous precautions, take also social distancing measures as a response to the epidemic. For example, they take time off work, cancel or postpone social events, or use less public transportation.
Indeed, the average number of behaviors changed by individuals who report at least one social distancing measure is much higher than the average of those who take only moderate preventive measures (respectively $6.26$ and $2.52$). This can be observed also in Fig~\ref{fig:behaviors}, where we show the histograms of number of behaviors changed in the two classes. Furthermore, individuals who report at least one social distancing measure do also report at least one moderate preventive measure. These observations support the idea that in our dataset are present at least two main forms of behavioral change: moderate preventive measure and social distancing. The latter can be regarded as a reinforcement of the former. Furthermore, the approach of dividing individuals into three classes aims at providing a more composite representation of behaviors. In \cite{reviewpsico} is underlined that most of current models do not allow for heterogeneous behavioral responses to an epidemic. However, this homogeneous assumption is broadly inconsistent with what we know about human behavior~\cite{abc,abc1}.

In summary, we divide individuals according to their responses into three classes:
\begin{itemize}
    \item individuals who do not change their behavior (defined as \textit{no change} in the following). This class corresponds to $26\%$ of responses;
    \item individuals who take only moderate preventive measures (defined as \textit{moderate change} in the following). This class corresponds to $36\%$ of responses;
    \item individuals who take also social distancing measures (defined as \textit{social distancing} in the following). This class corresponds to $38\%$ of responses.
\end{itemize}

Nevertheless, in the SI we report the results of the analyses done considering only two categories: 1) change 2) no-changes. The results show a classification performance significantly higher than the random benchmark, although the relative difference between the two is lower than with three categories. Most importantly, the most important features for classification in the two classes are consistent with those that emerge considering a tripartite division.

\subsection*{Features engineering}
As mentioned above, our dataset comprises intake, symptoms, and behavioral questionnaires collected from Influweb during the $2017-18$ and $2018-19$ flu seasons in Italy. We combine all these data obtaining $23$ features for each of the $599$ responses. In particular, the features have been created based on the following information: 1) socio-demographic indicators (age, gender, etc.), 2) health indicators (allergies, chronic diseases, frequency of flu episodes, etc.), 3) information indicators (whether the user actively sought information about the flu, self-assessment of the level of information about the disease), 4) feelings and beliefs towards the flu (concerns, impact on personal life of a possible contagion, etc.), 5) epidemic indicators (incidence of flu epidemics at the moment of response and timing of the peak respect to the moment of response).

In table \ref{tab:features} we provide a complete list of features with related meanings. Before moving forward, it is important to describe in some more details the construction of three key features. In particular, we define a \textit{disease score} that aims to provide a measure of the severity of the illnesses experienced in the past by participants. We define it as:

\begin{eqnarray}
\label{eq:diseasescore}
	\text{\textit{disease score}} = \sum_{i}{{M_{i}n_{i}e^{\gamma \Delta_{i}}}\over{n_{i}e^{\gamma \Delta_{i}}}}
\end{eqnarray}

Where $i$ runs over all the weekly (symptoms) questionnaires submitted by the individual; $M_{i}$ is the number of symptoms reported in the $i-th$ weekly questionnaire; $e^{\gamma \Delta_{i}}$ weighs the duration of illness giving more importance to the most recent ones. In particular, $\Delta_{i}$ is the difference - in years - between the submission date of the $i-th$ weekly and the behavioral survey, and $\gamma$ is a parameter that expresses how fast people forget past experiences (here set to $1$ year); $n_{i}$ is the number of individuals present in our dataset who reported their symptoms during the same week of the $i-th$ weekly questionnaire. Weighing observations with this term gives less importance to periods with just a few active participants and makes the \textit{disease score} of different individuals comparable. The \textit{disease score} can then be interpreted as follows: the higher, the more recent and severe the episode experienced by the individual. In the SI we test a much simpler definition of \textit{disease score} where we disregard the exponential temporal weights $e^{\gamma \Delta_{i}}$. Adopting a simpler definition slightly reduces the precision, but does not change the overall results. \\
We then define a \textit{perceived severity} measure. Participants are asked to evaluate some statements regarding the consequences of a possible contagion (for instance ``Flu would be a serious illness for me" or ``A contagion would have serious financial consequences for me"). This is done to asses the perceived impact that a possible contagion would have on individuals' life in general. Participants can express their level of agreement with statements through the possible answers probably true, probably false, do not know. We turn possible answers in numeric values, respectively $+1$, $-1$, and $0$ and we sum all answers to obtain the feature \textit{perceived severity}. It can assume values between $-4$ (minimum perceived severity) and $+4$ (maximum perceived severity).\\ 
Finally, we define a \textit{perceived susceptibility} measure. Participants are asked some questions regarding their feelings and perceptions towards the flu. The reduced STAI (State-Trait Anxiety Inventory) test is used as guideline for this questionnaire~\cite{marteau1992development,rubin2014design}. The goal is to assess their level of anxiety towards a possible contagion. To each of the questions they can answer either yes, no, do not know. We turn possible answers in numeric values, respectively $+1$, $-1$, $0$. Then, we sum all the answers. The resulting variable can assume values between $-4$ (minimum level of anxiety) and $+4$ (maximum level of anxiety). This to quantify the fact that individuals who are more anxious at the idea of becoming infected, also perceive themselves as more susceptible to the disease. 

It is important to notice how several features have been designed to match the constructs of the Health Belief Model (HBM)~\cite{hbm1,hbm2,hbm3}, which is by far the most commonly used psychological theory to explain and predict health-related behaviors. The underlying concept of HBM is that health behavior is determined by personal beliefs and perceptions about the disease: the more an individual feels \textit{threatened} by the possibility of infection, the more she will be inclined to embrace protective behaviors. More in detail, according to the HBM the \textit{perceived threat} of an individual is determined by two main constructs: 1) \textit{Perceived severity} refers to the individual's belief about the severity of the disease. The HBM proposes that individuals who perceive the disease as more severe are also more keen on protecting themselves through proactive behavioral measures. Even if \textit{perceived severity} is often based on medical information, it can also be a consequence of the beliefs a person has about the difficulties a disease would create on her life in general; 2) \textit{Perceived susceptibility}, instead, refers to the personal evaluation of the risk of contracting the disease. According to the HBM, individuals who consider themselves more vulnerable to the disease, are also much more likely to engage health-promoting behaviors. Of course, many other factors can influence individuals' decision-making process. In particular, the HBM suggests the existence of \textit{modifying variables} (such as socio-demographic indicators)  to explain interpersonal variability, and of endogenous events - called \textit{cues to action} - that prompt individuals towards the acceptance of healthier behaviors. The individual should also consider that behavioral change is actually decisive to decrease risk of contagion, and that the benefits associated with change are higher than the costs.

\begin{table*}
  \caption{List of features.}
  \label{tab:features}
  \resizebox{\columnwidth}{!}{
  \begin{tabular}{r|l|l}
    \hline
    \textbf{Type}&\textbf{Feature}&\textbf{Meaning}\\
    \hline
    \textbf{socio-demographic}&\textit{gender}&gender\\
    &\textit{age}&age class (15, 15-30, 30-50, 50-65, 65+ years)\\
    &\textit{contacts}&true if the individual has daily contacts with large groups, patients, children\\
    &\textit{smoke}&true if the individual smokes regularly\\
    &\textit{diet}&true if the individual follows a special diet\\
    &\textit{children}&true if the individual has children in school age\\
    &\textit{public transport}&true if the individual takes regularly public transportation\\
    &\textit{elderly}&true if the individual has old people (65+) in her household\\
    \hline
    \textbf{health-related}&\textit{flu frequency}& frequency of flu-like illness\\
    &\textit{flu}&true if the individual had flu in the current season\\
    &\textit{disease score}&measure of severity of diseases experienced in the past\\
    &\textit{vaccination}&true if the individual has received a vaccine in the current season\\
    &\textit{vaccination last year}&true if the individual has received a vaccine in the previous season\\
    &\textit{allergy}&true if the individual has allergies that can cause respiratory problems\\
    &\textit{disease}&true if the individual receives regularly medication for chronic diseases\\
    \hline
    \textbf{information-related}&\textit{info seeking}&true if the individual seeks regularly information regarding the flu\\
    &\textit{information}&self-evaluation of the level of information regarding the flu\\
    \hline
    \textbf{beliefs}&\textit{preventive}&true if the individual thinks that proactive measures can prevent the contagion\\
    &\textit{perceived susceptibility}&measure of anxiety deriving from a possible contagion\\
    &\textit{efficacy}& measure of awareness of efficacy of preventive measures\\
    &\textit{perceived severity}&measure of concerns related to the possibility of contagion\\
    \hline
    \textbf{epidemic indicators}&\textit{peak}&days between the ILI peak and the date of compilation of the behavioral survey\\
    &\textit{prevalence}&flu prevalence in the Italian region where the participant reside in\\
  \hline
\end{tabular}
}
\end{table*}

\subsection{Classification algorithms}

The data analyses are conducted with a range of machine learning algorithms. Decision tree ensembles are a powerful tool for classification tasks~\cite{decisiontree}. They consist in a set of classification trees. In fact, the underlying idea is that summing together the predictions of multiple ``weak" learners, one can achieve more robust predictions than with a single ``strong" learner. This general model is implemented by a great variety of algorithms, such as Gradient Boosted Trees (GBT)~\cite{greedy}. GBT exploits a specific training strategy called \textit{additive training}, in which at each training step is added to the ensemble the tree that optimizes the objective function. In this work, we use $XGBoost$, an open-source software library~\cite{xgboost, xgboostlib}. Recently, it has gained a great popularity for its speed and performance, and has become the algorithm of choice for many machine learning applications. In practice, we use this algorithm to classify individuals according to their features in the three behavioral change classes. To quantify the quality of the predictions of the GBT model we compare it to: i) a dummy classifier that generates random predictions by respecting the training set's class distribution. This is done to assess if and to which extent the GBT model performs better than a null benchmark, ii) other standard machine learning models such as Support Vector Machine (SVM)~\cite{svm}, Logistic Regression (LG) \cite{logisticregression} and Random Forest (RF)~\cite{randomforest}. We use the \textit{scikit-learn}~\cite{scikit-learn} implementation of these algorithms and we train them fine-tuning standard parameters. We refer the interested reader to the SI for more details and the code.

\subsection{Explainability}
The ultimate goal of our analysis is to determine which are the main drivers of behavioral change in response to epidemics and to which extent they influence people's behavior. To achieve this, understanding the hidden patterns spotted by the machine learning classifier is essential. To interpret model's decisions we exploit SHAP (SHapley Additive exPlanations), a unified approach that connects cooperative game theory with local explanations to explain the output of any machine learning model~\cite{shap1,shap2,shap3}. It aims at understanding the role and the significance of each feature in the model's decisions using Shapley values. The Shapley value is a solution concept in cooperative game theory that addresses the following issue: how important is each player to the overall cooperation, and what payoff can she reasonably expect? This is very similar to the problem we are considering. In our framework, the overall cooperation is the classification task, players are the features, and the payoff of players is the importance of features for the classification performance. We refer the readers to the SI for more details.

\section*{Results and discussions}

\subsection{Classification task}
After having pre-processed data and built the features, the next step of the analysis consists in training the models to classify individuals in the three classes of behavioral change.
Following a common approach, we divide our dataset in training set ($70\%$) and test set ($30\%$). Only the training set is used to find the optimal parameters for the models, while the test set is retained to evaluate performance. We search for optimal parameters using \textit{10-fold cross validation} over an extensive grid of candidate values. We use four different metrics (\textit{precision}, \textit{balanced accuracy}, \textit{recall} and \textit{f1 score}) to obtain a complete overview of the performance of the classification algorithm. From results in table~\ref{tab:gbt} we observe that GBT outperforms (across the four metrics) i) the trivial prediction strategy (RND), ii) other standard machine learning algorithms (SVM, RF, LG). It is important to notice how the highest precision obtained is far from $1$. However, to the best of our knowledge, this is the first paper taking such approach, thus we have no previous results (i.e. benchmarks) to compare and contrast with.  

\begin{table*}
  \caption{Classification performance..}
  \label{tab:gbt}
  \begin{tabular}{|c|c|c|c|c|}
    \hline
    \bf model&\bf precision&\bf bal. accuracy&\bf recall&\bf f1 score \\ \hline
    RND&0.343&0.335&0.334&0.335\\ \hline
    SVM&0.519&0.503&0.500&0.504\\ \hline
    LG&0.479&0.492&0.478&0.472\\ \hline
    RF&0.506&0.498&0.506&0.505\\ \hline
    GBT&\bf 0.546&\bf 0.549&\bf 0.550&\bf 0.546\\ \hline
    \end{tabular}
\end{table*}

\begin{figure}[t!]
  \centering
  \includegraphics[width=0.4\linewidth]{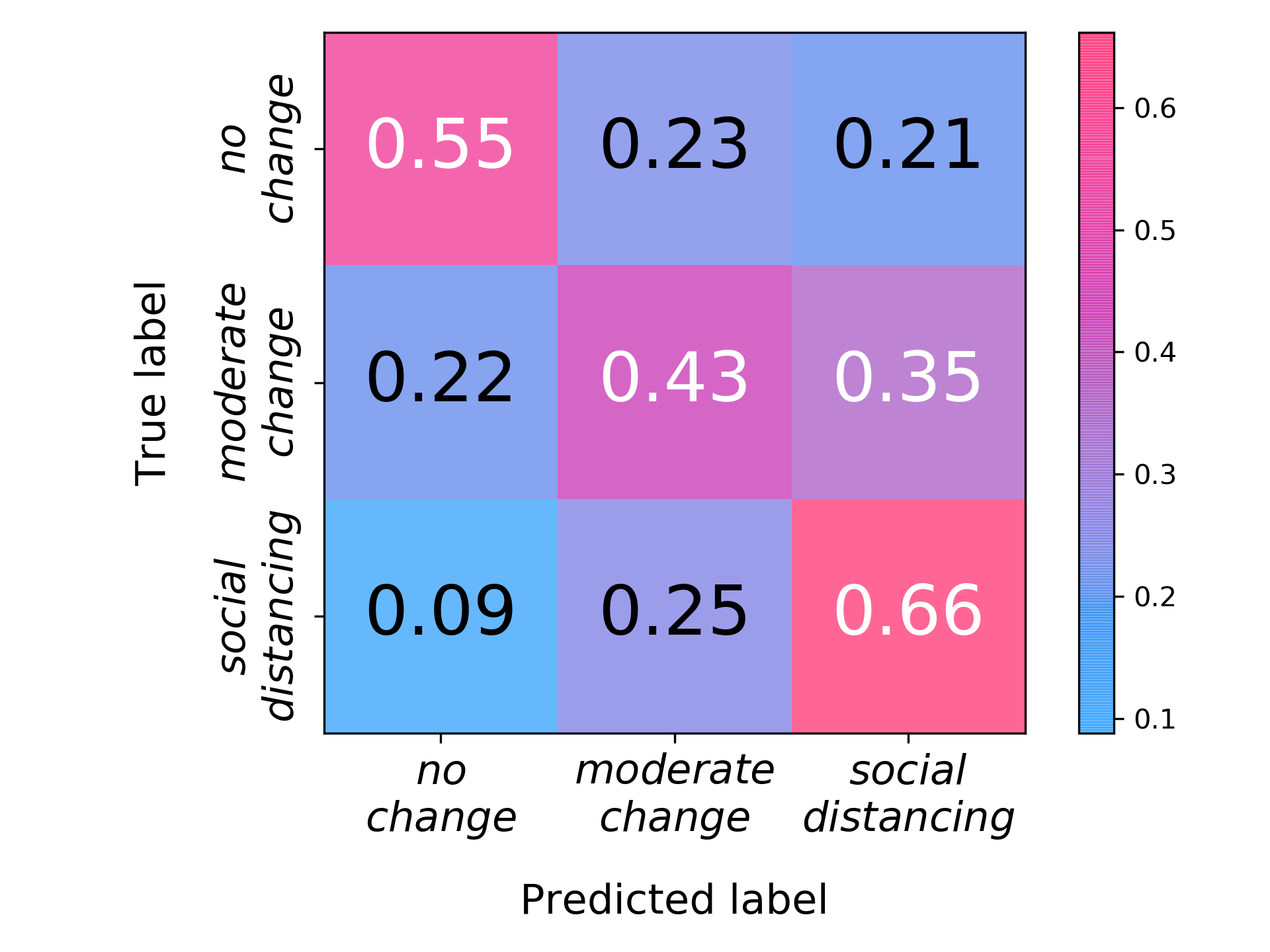}
  \caption{\textbf{Confusion matrix of GBT}. Each row and column represents a particular class: on the vertical axis are represented the true labels, while on the horizontal axis are represented the predicted labels. Hence, in the main diagonal boxes, we can observe the percentage of samples correctly labeled for each class, and in non-diagonal boxes, we can observe the percentage of misclassifications among all possible pairs of classes}
  \label{fig:confusion}
\end{figure}

Next, we analyze in depth the performance of GBT. Figure~\ref{fig:confusion} represents the confusion matrix of model's predictions. 
In statistical classification problems, the confusion matrix is a specific table layout that allows visualization of the performance of an algorithm. Interestingly, the best-predicted class is social distancing. In fact, $66\%$ of its samples are classified correctly. This result suggests that individuals changing their behaviors significantly stands out more in the feature space. Furthermore, most of the classification errors are between the two classes of behavioral change, while there are fewer errors between the two classes linked to changes in behaviors and the no change class. Thus, as one would expect, there is more similarity between responses in the two behavioral change classes than between responses of a behavioral change class and those of no change class.

\subsection*{Understanding model's decisions}
\label{sec:shap}
In this section we want to make a step further and inspect what GBT learns from data using SHAP, the tool of explainable machine learning that we have previously introduced.

\begin{figure}[t!]
  \centering
  \includegraphics[width=0.5\linewidth]{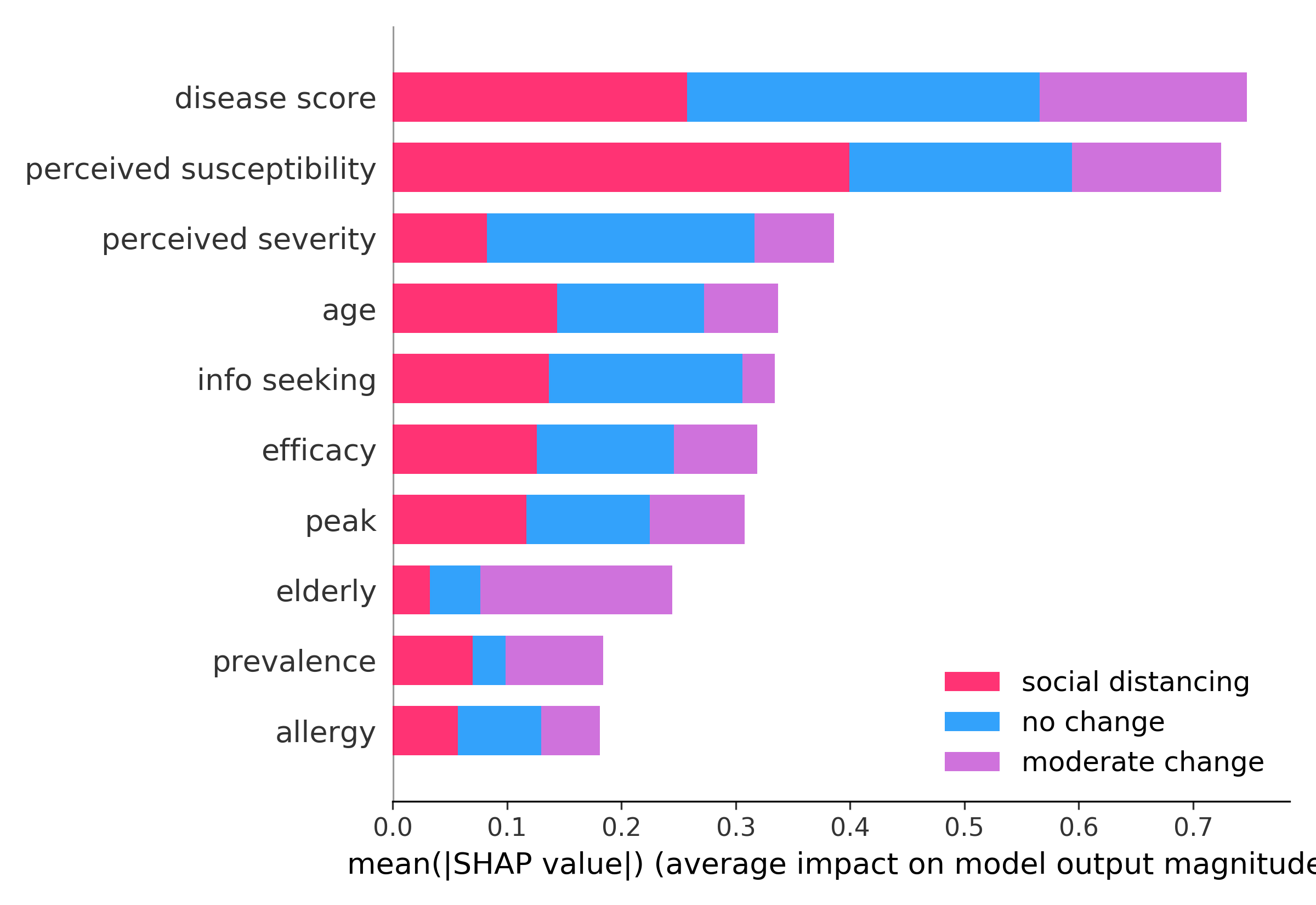}
  \caption{\textbf{Summary plot for SHAP analysis}. It shows the mean absolute SHAP value of ten most important features for the three classes}
  \label{fig:shapsummary}
\end{figure}

In Figure~\ref{fig:shapsummary} we have reported the mean absolute SHAP value of the ten most important features with respect to the three behavioral classes. This provides a general overview of the most influential features for the model and their impact on the classification of each behavioral class. Among the most determinant features we can recognize i) health-related factors (\textit{disease score}, \textit{allergy}) ii) personal beliefs (\textit{perceived susceptibility}, \textit{perceived severity}, \textit{efficacy}) iii) socio-demographic indicators (\textit{age}, \textit{elderly}, \textit{info seeking}), iv) information regarding the flu season (\textit{peak}, \textit{prevalence}). More in details, the top three features capture the severity and recency of past illnesses, the perceived susceptibility, and the perceived severity. Thus, having a history of illnesses induces users to be more careful and adapt their behaviors significantly. The last two features nicely match the HBM constructs associated to the drivers of behavioral changes. Notably, in seventh position we find the distance between the moment of response to the survey and the position of the peak. This suggests that the progression of the disease influences behaviors. Furthermore, seeking information about the flu has an important impact on the classification of class of no change and of social distancing. Thus, consulting news about the flu affects individuals' decisions. Finally, it is interesting to notice how living with elderly people is significantly linked to moderate changes in behaviors. Users conscious of the risks of infection for elderly individuals might modify their behaviors as preventive measure. This highlights how behavioral changes are indeed a complex phenomenon possibly driven, at least in part, by altruistic concerns for others. 

\begin{figure}[t!]
  \centering
  \includegraphics[width=0.6\linewidth]{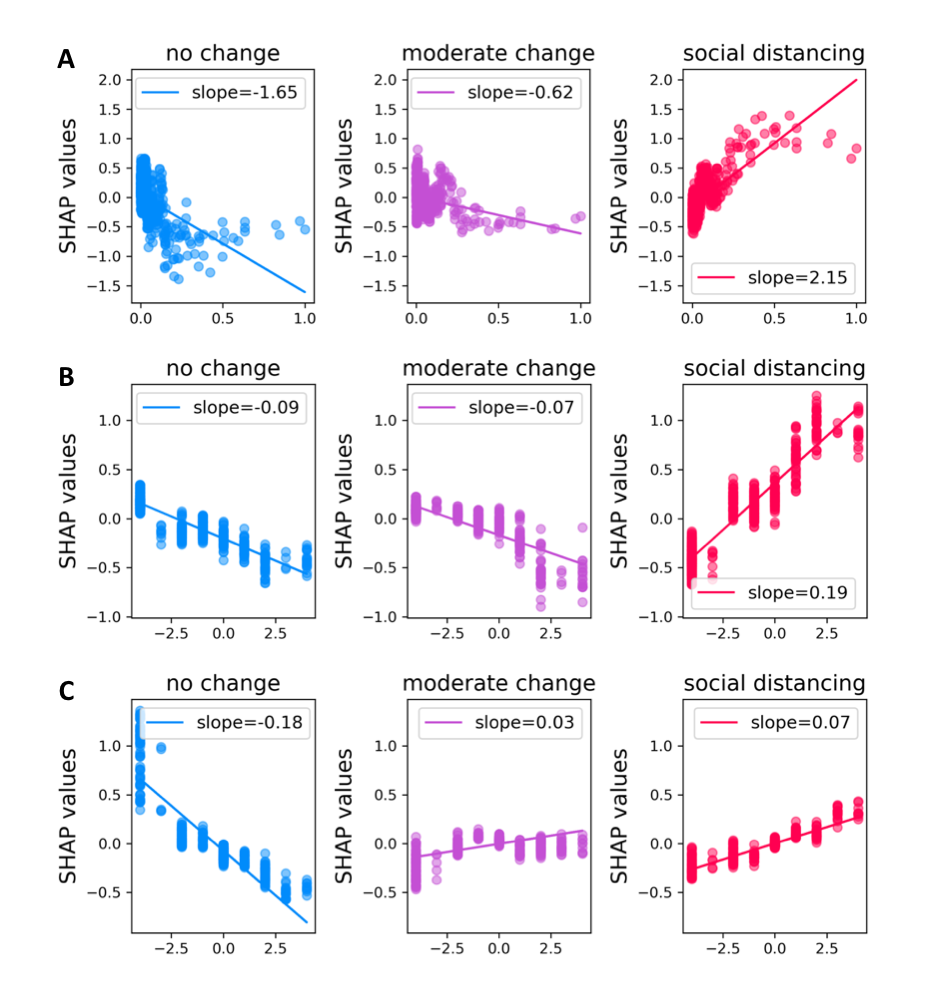}
  \caption{SHAP value plot for the three most important features:  A) \textit{disease score}, B) \textit{perceived susceptibility}, C) \textit{perceived severity}.}
  \label{fig:shapresults}
\end{figure}

In order to deepen our understanding, in Fig \ref{fig:shapresults} we study the relation between SHAP and features values for the three most important features: \textit{disease score}, \textit{perceived susceptibility}, and \textit{perceived severity}. In particular, we plot the scatter plots in which the x-axes describe the feature and the y-axes the SHAP value for each of the $599$ responses. In Fig \ref{fig:shapresults}A we inspect the effect of disease score on model's decisions. We observe that having higher disease score has a huge positive effect on distancing measures. We observe that ``low" and ``medium" values of disease score have a positive non-negligible effect on the probability of adopting moderate behavioral measures. This observation is important to stress how a single variable might not be enough to capture the adoption of a particular behavioral category. In Fig \ref{fig:shapresults}B, we can observe that for the class of social distancing, SHAP values increase from negative to positive values as a function of \textit{perceived susceptibility}. This suggests that those who perceive themselves as more susceptible to a possible contagion are more likely to adopt social distancing measures. Reasonably, for class of no change the trend of SHAP values is decreasing, meaning that those who do not feel susceptible will not probably change their behavior. The effects for the class of moderate change are similar to that of no change just discussed, even if we observe a weaker downward trend. In Fig \ref{fig:shapresults}C, we can observe that for the class of social distancing SHAP values increase from negative to positive values as a function of \textit{perceived severity}. This prompts us to conclude that individuals who perceive the disease as more severe are also more likely to protect themselves through the adoption of social distancing measures. On the other hand, since for class of no change SHAP values show a decreasing trend, we can conclude also that those who perceive the disease as not particularly severe will not probably change any of their behaviors. 

Overall, these observations are in very good accordance with what is assumed in the HBM regarding the existence of beliefs constructs - such as \textit{perceived susceptibility} and \textit{perceived severity} - that influence the probability of adopting safer behaviors. We can thus conclude that, not only the GBT model independently selects as fundamental drivers of behavioral change the belief constructs suggested in the HBM, but also that their effect is the same as theorized in the HBM. Furthermore, the result highlights the importance of personal past experiences with illnesses.
From a public health perspective this suggests that, in order for communication campaigns
 to be more effective, they should leverage the individuals' personal experience and stress how recommended behaviors might help achieve more positive personal health outcomes. 

\section*{Conclusion}
Our understanding of behavioral changes induced by infectious diseases is unfortunately extremely limited and anecdotal. The lack of data isolating and capturing this complex phenomenon is the key challenge. Here, we studied a unique dataset comprised of $599$ responses to a questionnaire about behavioral changes submitted by $N=434$ volunteers of the participatory Web platform Influweb during the $2017-18$ and the $2018-19$ flu seasons in Italy. 
For each response, we identified $23$ features regarding socio-demographic information, personal history of illnesses, and sentiment about the flu epidemic, and one target class describing the type of behavioral change implemented by respondents. Then, we investigated the possibility of predicting these target classes from the features adopting a range of machine learning algorithms. 
 Gradient Boosted Trees outperformed a random predictor, \textit{SVM}, \textit{Random Forest}, and \textit{Logistic Regression}. While the average precision (across the three classes) of the best model is only $0.546$, $66\%$ of the samples belonging to the class of most drastic behavioral changes were correctly identified. It is important to notice how, to the best of our knowledge, there are not similar studies to compare and contrast our results. In fact, as mentioned above, the study of behavioral changes induced by disease has been mostly a theoretical endeavour.

Since we are interested in understanding the factors driving people to change behaviors, we investigated the patterns spotted by the GBT model. To this end, we exploited a recent tool in the field of explainable machine learning: SHAP. By using this approach we discovered that the intensity and the recency of past personal episodes of illnesses, perceived susceptibility and perceived severity of an infection are the most important features used for classification. These findings highlight the importance of negative past experiences and are in very good accordance with the expectations from the Health Belief Model. In fact, this theoretical framework predicates the existence of beliefs constructs (such as \textit{perceived susceptibility} and \textit{perceived severity}) as main drivers of behavioral change induced by epidemics. In the top ten of most important features we found also i) the timing of response in relation to the peak of the seasonal flu, ii) the extent to which participants sought information about the flu, iii) whether the participant was living with elderly people. These results suggest that indeed the progression of the season induces changes in behavior, that seeking information about the disease might affect individuals' decisions and that individuals might change behaviors as a form of altruistic protection. 
In the SI, we verified the robustness and validity of our results by training the model on different subsets of our dataset and changing the definition of a key variable.

The presented study comes with limitations. First, while the socio-demographic indicators of participants are in line with the Italian population as a whole (see SI), the sample might be affected by self-selection biases. Indeed, we queried users willing to devote their time to the monitoring of the seasonal flu. Their sentiments, concerns and thus behaviors in response to the disease might deviate from those of the general population. Second, although, there are no other studies to compare the precision of our prediction task with, the absolute value of it is satisfactory, at best. The analysis of SHAP values and the connections of our findings with the well known theoretical constructs of the HBM are definitely reassuring. Nevertheless, future work is needed to collect larger samples and to quantify the general validity of these results. Third, we identified three classes of behavioral changes. Arguably, this classification could be refined to account for much more heterogeneity. To this end, the collection of larger samples is key. Finally, the study focuses only on one piece of the puzzle. Indeed, we did not investigate the effects of behavioral changes on the unfolding of the disease. Behaviors and diseases are linked by a feedback loop. Here, we simply focused on the first. Challenging future work is needed to connect these observations with the disease dynamics and back to the behaviors.\\
Overall, the research is a step towards the characterization of the factors driving behavioral changes during an outbreak. The study contributes to the small body of empirical literature on the subject and paves the way to future extensions and generalizations necessary to improve our understanding of human adaptive behaviors. From a public health stand point, a data-driven characterization of the key factors influencing changes in behaviors opens new perspectives in the possibility of devising more effective communication campaigns aimed at mitigating flu transmission among individuals.

\begin{acknowledgments}
Nicol\`o Gozzi thanks the Doctoral Training Alliance (DTA) and the Marie Sk\l odowska-Curie Fellowship programme for the possibility of taking part and receive funding through the DTA3/COFUND PhD programme. Authors thank also Dr Vittoria Colizza and Dr Alain Barrat for helpful discussions and suggestions.
\end{acknowledgments}

\newpage
\appendix
\section{Appendix}

\subsection*{Descriptive analysis of data}
We report here a few qualitative observations on the data. First, responses are divided in the three behavioral categories as follows: $26\%$ report no changes, ii) $36\%$ only moderate changes, iii) $38\%$ significant variations. Thus the majority of them describe either moderate or severe changes in behaviors. 

Second, only $20\%$ of the participants that reported variations in their behaviors in a particular season, experienced the flu in that season before or while compiling the survey. We identify the cases of flu using the definition of the European Centre for Disease Prevention and Control (ECDC)~\cite{ecdcdefinition}. According to this definition, seasonal flu is identified by the following criteria: sudden onset of symptoms, at least one systemic symptom (fever or feverishness, malaise, headache, myalgia), and at least one of respiratory symptom (cough, sore throat, shortness of breath).

\begin{figure}[t!]
  \centering
  \includegraphics[width=0.8\linewidth]{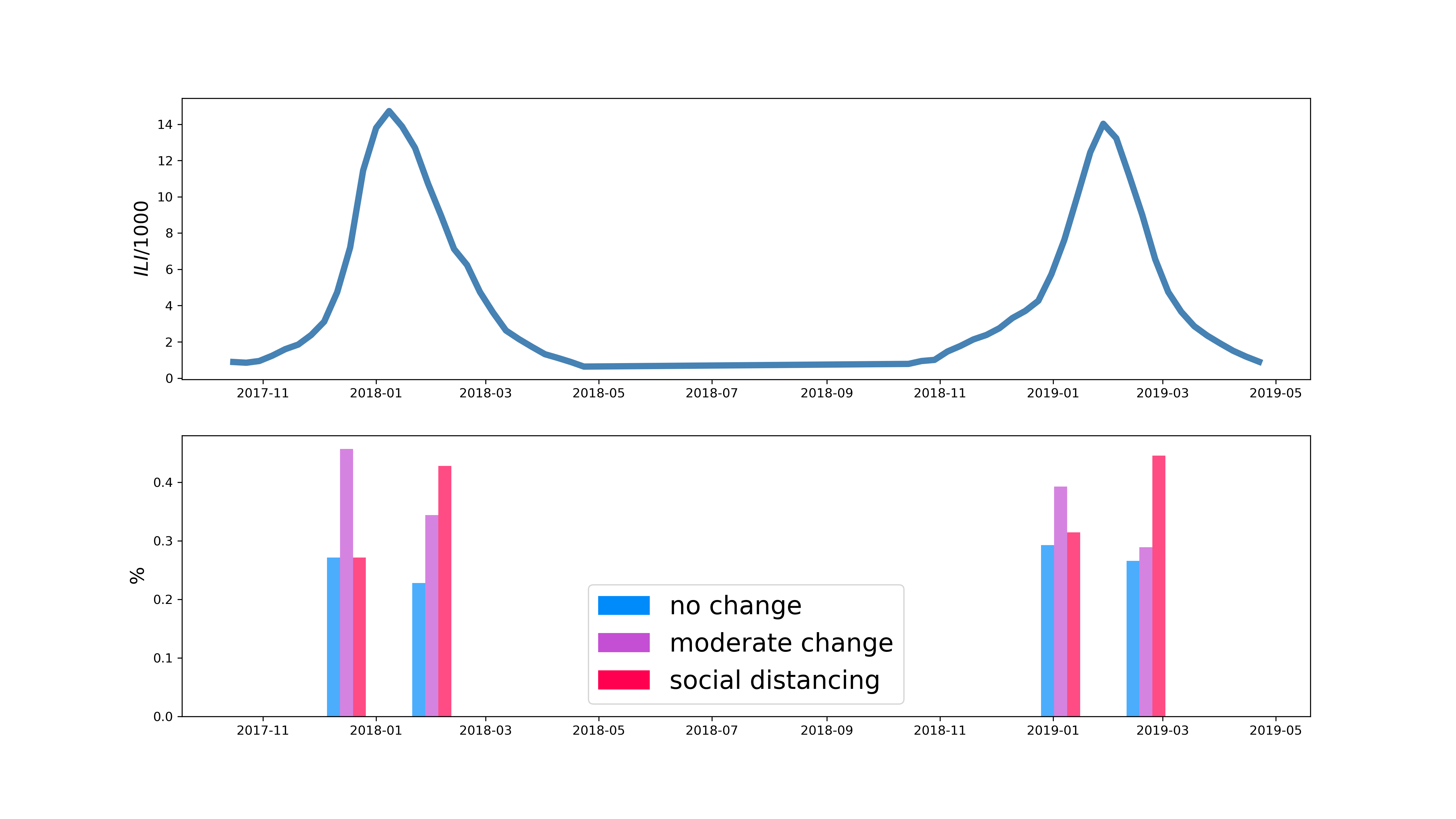}
  \caption{Flu incidence in the two seasons (above) and behavioral class distribution before/after the peak (below). The two seasons had a similar impact in terms of epidemic incidence, while the peaks are slightly shifted (2nd week of 2018 for the 2017/18 flu season, and 5th week of 2019 for the 2018/19).}
  \label{fig:fluseason}
\end{figure}

Third, during $2017-18$ season, $27\%$ of the answers provided before the peak are related to the category of no change, $46\%$ to moderate change and $27\%$ to social distancing. After the peak, instead, the number of surveys related to no change is $23\%$, to moderate change is $34\%$, while the percentage of surveys related to social distancing significantly grows up to $43\%$. Very similarly, during $2018-19$ season before the peak we have $29\%$ of surveys that report any change in behavior, $39\%$ that report only moderate change to behavior, and $32\%$ that report the adoption of social distancing measures. The division after the peak, instead, is the following: $26\%$ of surveys related to no change, $29\%$ to moderate change and $45\%$ to social distancing (Fig \ref{fig:fluseason}). Interestingly, in both seasons we note a significant increase in the number of surveys reporting social distancing measures after the peak. This observation highlights how risk perception is linked to the progression of the disease and not constant throughout. Indeed, proactive behavioral measures adopted by people seem to strengthen when external conditions get worse. 

\begin{figure}[t!]
  \centering
  \includegraphics[width=0.6\linewidth]{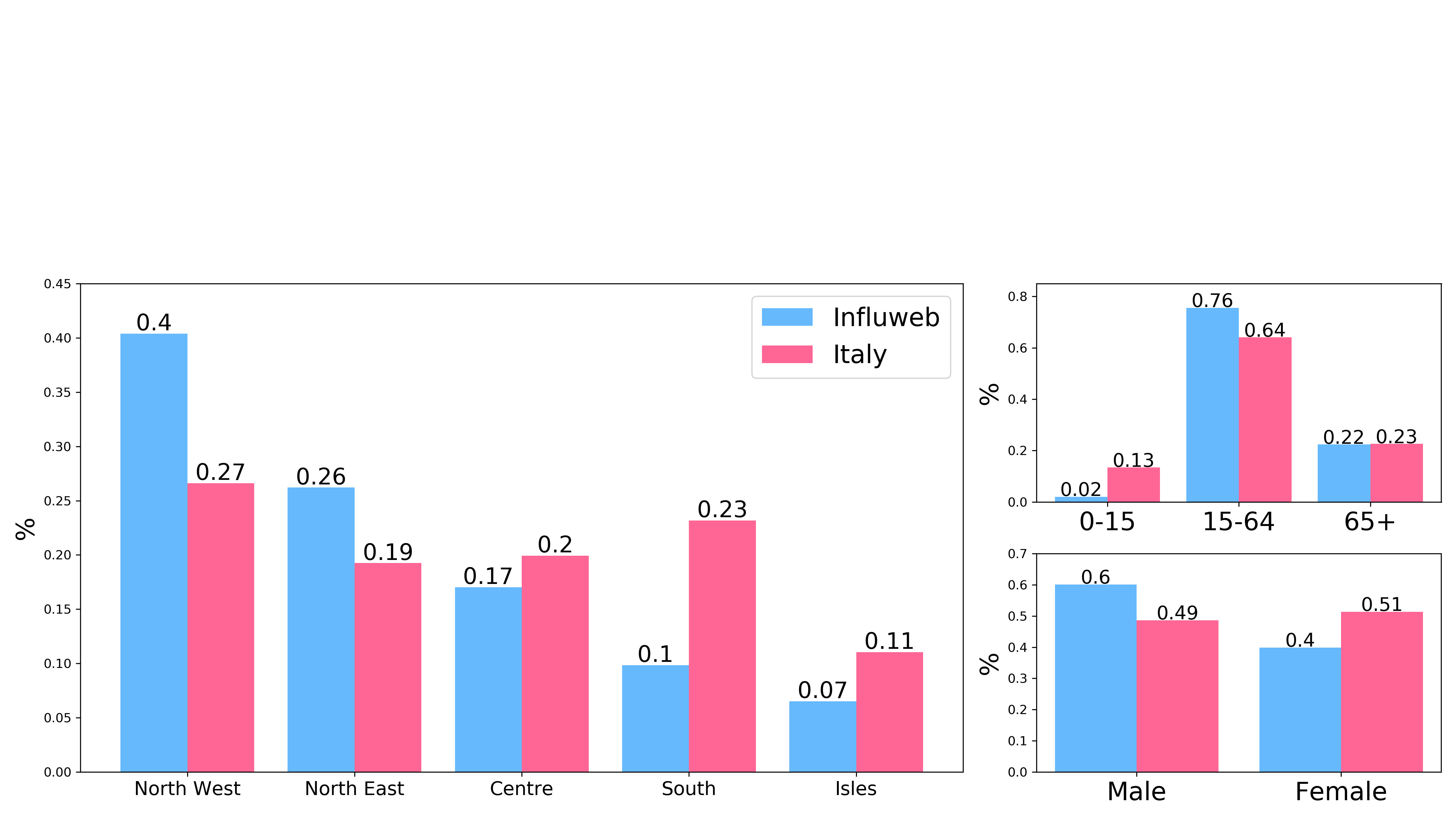}
  \caption{Comparison of our sample with the general Italian population in terms of age, gender and home area.}
  \label{fig:istat}
\end{figure}

Fourth, the comparison of our sample with the statistics retrieved from the online archive of the Italian National Institute of Statistics (ISTAT)~\cite{istat} in Fig \ref{fig:istat} shows that the geographic, gender and age distributions of participants are almost in line with the Italian population as a whole. In particular, we note a slight over representation of i) Northern Italy (+13\% north west, +7\% north east), ii) of male (+11\%), and iii) of age group 15-64 (+12\%).

\subsection*{Classification algorithms}
\textbf{Decision tree ensembles} are a powerful tool for classification tasks~\cite{decisiontree}. They consist in a set of classification trees. The underlying idea is that summing together the predictions of multiple ``weak" learners, one can achieve more robust predictions than with a single ``strong" learner. Mathematically, the prediction $\hat{y}_{i}$ of this model for the \textit{i-th} point of the dataset can be written as:
\begin{eqnarray}
\hat{y}_{i} = \sum_{k=1}^{K} f_{k}(x_{i})
\label{eq:decisiontree}
\end{eqnarray}
Where $K$ is the number of trees (i.e. the size of the ensembles) and $f_{k}(x_{i})$ is the prediction of the \textit{k-th} tree for the \textit{i-th} point. This general model is implemented by a great variety of algorithms, such as Gradient Boosted Trees (GBT)~\cite{greedy} and Random Forest (RF)~\cite{randomforest}. The differences arise from how trees are built and added. GBT exploits a specific training strategy called \textit{additive training}, in which at each training step is added to the ensemble the tree that optimizes the objective function. Recently, GBT has gained a great popularity for its speed and performance, and has become the algorithm of choice for many machine learning applications. 

\textbf{Support Vector Machines} (SVM)~\cite{svm} extends the concept of Linear Classification by trying to find a hyperplane in an N-dimensional space (where N is the number of features) that distinctly classifies the data points and that has the maximum margin. This approach is more stable to noise and to the classification of future points. However, one of the main advantages of SVM is the so-called ``kernel trick". In fact, using this method SVM can easily handle also seriously non-separable data. Through the kernel trick data can be easily projected in a vector space where they are - eventually - linearly separable. The only request for this new space is that the dot product (called \textit{kernel}) must be defined. Commonly used kernels are the polynomial kernel or the radial basis function kernel (RBF). This approach makes SVM a very powerful tool that can be exploited to classify a broad range of datasets. In this work we consider a SVM with RBF kernel. We tried also a SVM with linear kernel (results not shown) with comparable results.

\textbf{Logistic Regression} (LG)~\cite{logisticregression} is a very common tool used to predict probabilities. Given a sample input vector $x \in \mathbb{R}^{n}$ and the vector of model's weights $\theta \in \mathbb{R}^{n}$, LG smoothly projects the signal $s = \theta^{T}x$ in the probability range $[0,1]$ using - for example - a logistic function defined as: 
\begin{eqnarray}
\gamma(s) = \frac{e^{s}}{1+e^{s}}
\label{eq:logisticfunction}
\end{eqnarray}
The output can then be interpreted as a probability for a binary event. This framework can be easily extended to the cases with more than two possible outcome.

\textbf{Dummy Predictors} (RND) are generally used as null benchmarks to assess whether the models learnt something from data. There are many trivial prediction strategies, for example constant or equally probable predictions (``coin toss"). In this work we consider a random predictor that generates random predictions by respecting the training set's class distribution. 

For all the models we use open-source implementation. In particular, we use $XGBoost$, an open-source software library which provides a ``scalable, portable and distributed implementation" of GBT~\cite{xgboost, xgboostlib}. For other models we use the implementation provided by the open-source library \textit{scikit-learn}~\cite{scikit-learn} and we train them fine-tuning standard parameters. The source code can be find at \cite{githubcode}.

\subsection*{Sensitivity analysis}
We test here the stability of classification results of GBT. To this end, we train the model on various subsets of the whole dataset. First, we classify separately the surveys submitted during the $2017-18$ ($N=331$ surveys) and the $2018-19$ ($N=268$ surveys) flu season. We obtain a $precision$ of $0.572$ for the first and of $0.474$ for the second season. These results are still significantly higher than the random prediction and are comparable to those obtained in the general case with the two seasons aggregated ($precision=0.546$). Second, we repeat the classification excluding from the set of features those extracted from behavioral surveys. The goal of this analysis is to explore the possibility of inferring attitudes towards behavioral change on a much bigger scale. In fact, behavioral questionnaires represent an additional burden for the participants and mostly important Influweb is one of $9$ platforms in Europe. As reported in table \ref{tab:sensitivity}, disregarding features deriving from behavioral surveys leads to a $precision$ of $0.435$. Despite an expected decrease in model's performance, results are still better than the random guess. These observations are encouraging and shed light on the possibility of extending of our approach. However, larger samples as well as data from other countries are needed to reinforce and scale our findings. Third, we consider a much simpler definition of the most important feature used in classification: \textit{disease score}. In particular we consider the average number of symptoms reported by different individuals in symptoms questionnaires as a measure of severity of past illnesses. In other words, we disregard the exponential temporal weights of the original definition. Looking at results in table \ref{tab:sensitivity}, we note that the adoption of this simpler definition of \textit{disease score} does not significantly affect results, even if the classification performance across the four metrics is slightly lower. Fourth, we repeat the analysis considering only one survey for each user (we consider the most recent ones). In table \ref{tab:sensitivity} we can see that the classification performance is still higher than the random benchmark. Finally, we consider $2$ behavioral classes instead of $3$. The problem then become a binary classification between the class of \textit{no change} and that of \textit{any change}. We report a $precision$ of $0.743$ with respect to $0.618$ of the random prediction. Notably, the five most important features in the \textit{unique} and \textit{2 classes} cases are still those emerged in the complete 3-class problem (\textit{perceived susceptibility}, \textit{disease score}, \textit{info seeking}, \textit{perceived severity} and \textit{age}).  

\begin{table*}
  \caption{GBT sensitivity analysis.}
  \label{tab:sensitivity}
  \begin{tabular}{|c|c|c|c|c|}
    \hline
    \bf model&\bf precision&\bf bal. accuracy&\bf recall&\bf f1 score \\ \hline
    RND&0.343&0.335&0.334&0.335\\ \hline
    2017/18&0.572 &0.556 &0.550 &0.553 \\ \hline
    2018/19&0.474 &0.455 &0.444 &0.441 \\ \hline
    no behavioral survey&0.435 &0.432 &0.439 &0.435 \\ \hline
    simplified disease score&0.530 &0.533 &0.528 &0.524 \\ \hline
    unique&0.478 &0.461 &0.473 &0.472 \\ \hline
    2 classes&0.743(0.618) &0.646(0.504) &0.761(0.503) &0.748(0.532) \\ \hline
    \end{tabular}
\end{table*}

\subsection*{SHAP, an example}
In SHAP, the classic definition of the Shapley value is adapted to the specific problem of assigning to each feature its importance in the classification task. The result is called SHAP value and, for feature \textit{i}, is defined as:
\begin{eqnarray}
\phi_{i} = \sum_{S \subseteq N \setminus \{i\}} \underbrace{\frac{|S|! (M - |S|-1)!}{M!}}_\text{(1)}\underbrace{[p(S \cup \{i\})-p(S)]}_\text{(2)}
\label{eq:shapley}
\end{eqnarray}
Where, $S$ are all the possible subsets of features without $i$; $|S|$ is the number of features in $S$; $M$ is the total number of features; $p(S \cup \{i\})$ is the outcome of prediction considering both $i$ and features in $S$, and $p(S)$ is the outcome considering only features in $S$. To better understand the meaning of $\phi_{i}$, we divide its expression in blocks. In expression \ref{eq:shapley}, part (1) is an interaction term that accounts for all possible sequences in which features are added: in fact, the order in which features are taken into account is meaningful to how their importance is assigned. Given this, we move to part (2). Intuitively, if feature $i$ has a negligible impact on model's decisions, then considering it or not should not affect the prediction. Looking at \ref{eq:shapley} this is equivalent to saying: 
\begin{eqnarray}
p(S \cup \{i\}) \simeq p(S) \rightarrow \phi_{i} \simeq 0
\end{eqnarray}
On the other hand, if feature \textit{i} is decisive for the classification we expect $p(S \cup \{i\}) \neq p(S)$ and $\phi_{i} \neq 0$. As a guiding criterion, if feature \textit{i} is \textit{important} for the classification of an individual then its SHAP value will be significantly different from zero. More in detail, if it is negative, the feature under consideration lowers the likelihood of belonging to that particular class, if it is positive it raises it.

\begin{figure}[t!]
  \centering
  \includegraphics[width=0.6\linewidth]{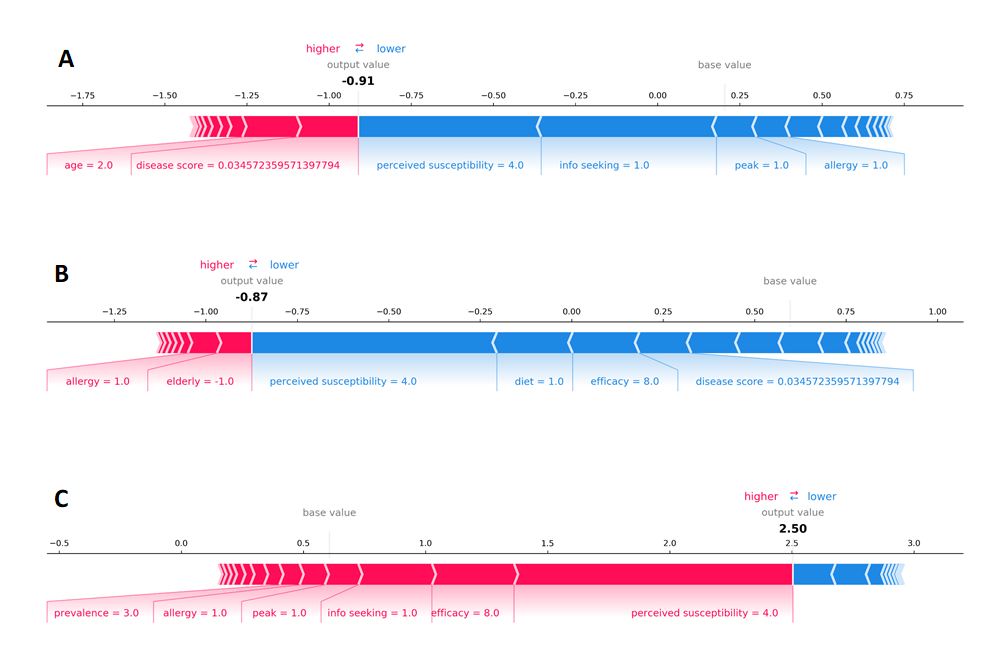}
  \caption{SHAP example.}
  \label{fig:shap}
\end{figure}

In this appendix we present also an example on how SHAP works and how it helps in understanding model's decisions. In  Fig~\ref{fig:shap} we try to explain the class prediction for a specific individual made by GBT. The chart is composed of three figures and each one refers to a specific class: \ref{fig:shap}A is related to the no change class, \ref{fig:shap}B to moderate change, and \ref{fig:shap}C to social distancing. The individual under examination belongs to the class of social distancing and is classified correctly. This can be figured out examining the output of the model for each class: $-0.91$ for no change class, $-0.87$ for moderate change, and $2.50$ for social distancing. Intuitively, these numbers indicate that the most unlikely class for this individual is no change. A little bit more likely is moderate change, but the class of social distancing is far more probable. However, SHAP provides us much more information. In fact, we can evaluate the role played by each feature in the classification: the blue ones in figure have a negative effect on the likelihood of belonging to that particular class, while the red ones have a positive effect. Furthermore, the length of the arrow (red or blue) related to each feature represents its SHAP value. Then, for example, we can conclude that for this individual, \textit{info seeking} and \textit{perceived susceptibility} both have an important negative impact on the likelihood of being in the no change class. Conversely, they have a great positive impact on the likelihood of being in the correct class, social distancing. Finally, we can give an interpretation of the prediction: since the individual feels vulnerable to the disease (\textit{perceived susceptibility}$=4$) and searches regularly for information regarding the flu (\textit{seek}$=1$), then she is with great confidence a member of the class of social distancing.

\newpage
\providecommand{\noopsort}[1]{}\providecommand{\singleletter}[1]{#1}%

\end{document}